# Petri Nets and Machines of Things That Flow

Sabah Al-Fedaghi
Computer Engineering Department
Kuwait University
Kuwait
sabah.alfedaghi@ku.edu.kw

Dana Shbeeb
Projects Implementation Department
Public Authority for Sports
Kuwait
dana_shbeeb@hotmail.com

*Abstract*—Petri nets are an established graphical formalism for modeling and analyzing the behavior of systems. An important consideration of the value of Petri nets is their use in describing both the syntax and semantics of modeling formalisms. Describing a modeling notation in terms of a formal technique such as Petri nets provides a way to minimize ambiguity. Accordingly, it is imperative to develop a deep and diverse understanding of Petri nets. This paper is directed toward a new, but preliminary, exploration of the semantics of such an important tool. Specifically, the concern in this paper is with the semantics of Petri nets interpreted in a modeling language based on the notion of machines of "things that flow." The semantics of several Petri net diagrams are analyzed in terms of "flow of things." The results point to the viability of the approach for exploring the underlying assumptions of Petri nets.

I. INTRODUCTION

Petri nets are an established graphical formalism for modeling and analyzing the behavior of systems, including hardware and software systems, communication systems, and manufacturing systems [1]. The classical Petri net is defined as a directed bipartite graph with two types of nodes: places (i.e., conditions) and transitions (i.e., events that may occur). Connections between two nodes of the same type are not allowed. Places are represented by circles and transitions by rectangles [2].

As of 2004, more than 8,000 publications on Petri nets (see http://www.informatik.uni-hamburg.de/TGI/PetriNets/) have been produced since their introduction in 1962. Petri nets have been utilized to support UML notations and their transformations [3]. They have also been applied in networks requirements analysis and performance estimation, computer hardware architecture modeling, architectural specification for distributed systems, system on a chip verification, real time systems critical performance estimation, fault diagnosis, workflow analysis, traffic control, and e-commerce system modeling [3]. "Compared with other formalisms Petri nets are preferable for visualization and comprehension by different stakeholders" [3].

This graphical representation [of Petri nets] is very simple, intuitive, and very appealing, even for people who are not very familiar with the formal details. With only three kinds of graphical items (circles, boxes, and arrows) it is possible to mostly describe any kind of process and to obtain, for free, a very clear and qualitative conceptual model of process itself. Thus, nor a strong knowledge of mathematics, neither a good knowledge of any programming language is required, and this may represent a consistent leap ahead in respect to classical approaches, since it could allow to life-scientists to directly use with little or no effort at all. [4]

Nevertheless, others disagree with such claims and consider Petri nets quite technical for nontechnical persons to understand and read. For example, in business processes, BPMN [5] is "a better candidate to visualize more complicated processes, which is better looking and more descriptive for business people" [6].

A. Motivation

The extensive use of Petri nets has given rise to different schools of thought concerning the semantical interpretation of nets, with each view justified either by the theoretical characterization of different properties of the modelled systems, or by the architecture of possible implementations. [7]

An important consideration in the value of Petri nets is their utilization for describing both the syntax and semantics of the modeling formalism. Modeling notations have been criticized as imprecise. Describing a modeling notation in terms of a formal technique provides a way to minimize ambiguity. Suitable techniques for doing so include general-purpose modeling approaches such as Petri nets [8].

*Accordingly, it is imperative to develop a deep and diverse understanding of Petri nets. The present paper is directed toward a new, but preliminary, exploration of the semantics of such an important tool.*

B. Approach

This paper is concerned with Petri nets in the area of diagrammatic conceptual modeling. According to Thalheim [9], a model is a representation of an aspect of the real world with aims that include (1) facilitating understanding by eliminating unnecessary components, (2) aiding in decision making, (3) explaining, controlling, and predicting events in a system. There are other important objectives such as development and design processes. Specifically, the concern here is with the semantics of Petri nets interpreted in terms of a modeling language based on the notion of "things that flow" (to be defined later). We assume a basic knowledge of the concept of Petri nets.

For the sake of self-contained paper, the flowthing machine (FM) model will be reviewed briefly in the next section. FM has been utilized as a modeling tool in several fields, including software engineering, business processes, and engineering design [10-12].

## II. FLOWTHING MACHINE MODEL

Diagrammatic modeling is needed at the level between "natural communication" (e.g., spoken language) and semiformal specification to provide a systematic way of thinking about events in the theater of operations. This paper utilizes a diagrammatic language that depicts *machines* comprising five basic "operations": creating, releasing, transferring, receiving, and processing of *things*. This type of language can play a central role in facilitating understanding among all participants and as a first step toward developing and facilitating policies and implementation plans.

An abstract machine is a diagrammatic schema that uses *flow things* to represent all types of physical and non-physical entities. Flowthings *flow* among basic machine stages in which a flowthing can be created, released, transferred, processed, and received (see Fig. 1). Hereafter, flowthings may be referred to as *things* and an abstract flow machine as a *machine*.

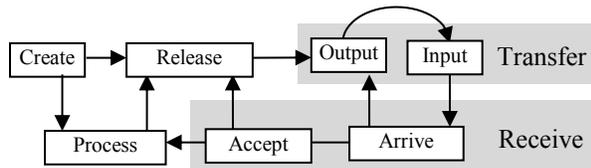

Fig. 1. Flow Machine

The machine is the conceptual structure used to change or transmit things as they pass through stages, from their inception or arrival to their de-creation or transmission. Machines form the organizational structure (blueprint) of any security system. These machines can be embedded in a network of assemblies called *spheres* in which the machines operate. The stages in Fig. 1 can be described as follows:

**Arrive**: A thing reaches a new machine.

**Accepted**: A thing is permitted to enter, or not (e.g., missing or wrong credentials). If arriving things are always accepted, *Arrive* and *Accept* can be combined as a **Received** stage.

**Processed** (changed): A thing goes through some kind of transformation that changes it without creating a new thing (e.g., a traveler is processed at a passport checkpoint).

**Released**: A thing is marked as ready to be transferred outside the machine (e.g., a passenger is cleared to enter the boarding area).

**Transferred**: A thing is transported somewhere from/to outside the machine (from one airport to another).

**Created**: A new thing appears in a machine (e.g., a search exposes a weapon among the things flowing in the search machine)

The machine shown in Fig. 1 is a generalization of the typical *input-process-output* model used in many scientific fields. The stages in this machine are mutually exclusive. An additional stage of *Storage* can be added to any machine to represent the storage of things, but storage is not an exclusive stage because there can be *stored processed* flowthings, *stored created* flowthings, etc.

The notion of *spheres and subspheres* refers to network environments (e.g., the SSCP machine is within the *sphere of the arrival terminal*). Multiple machines can exist in a sphere if needed. The machine is a subsphere that embodies the flow; it itself has no sub-spheres. *Triggering* is the activation of a flow, denoted by a dashed arrow. It is a dependency among flows and parts of flows. A flow is said to be triggered if it is created or activated by another flow (e.g., the exit-flow from a queue triggers an in-flow of waiting passengers). Triggering can also be used to initiate events such as starting up a machine (e.g., a manager's signal triggers the opening of an additional queue to alleviate overcrowding).

## III. EXAMPLE

MIT OpenCourseWare [13] presents a state transition diagram for a parking gate controller, shown in Fig. 2. "The machine has four possible states: 'waiting' (for a car to arrive at the gate), 'raising' (the arm), 'raised' (the arm is at the top position and we're waiting for the car to drive through the gate), and 'lowering' (the arm)." Fig. 3 shows a suggested implementation.

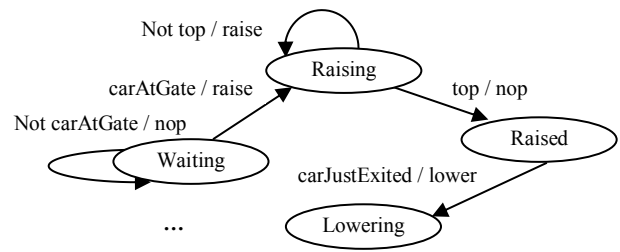

Fig. 2. State diagram for parking gate controller (redrawn, partial from [14])

```
if state == 'waiting' and carAtGate:
    nextState = 'raising'
elif state == 'raising' and gatePosition == 'top':
    nextState = 'raised'
elif state == 'raised' and carJustExited:
    nextState = 'lowering'
elif state == 'lowering' and gatePosition == 'bottom':
    nextState = 'waiting'
else:
    nextState = state
```

Fig. 3. Partial view of the implementation of parking gate controller (redrawn, partial from [13])

*A. The Static Description*

Fig. 4 shows a possible FM representation of the parking gate controller. When a car is received in the area just before the arm (circle 1) and the arm is not in the top position, then this triggers (2) transferring the arm to the top position (3 and 4). The car then proceeds to the arm area (5 and 6). When the car is received in the area just after the arm (7), this triggers (8) the arm to change position to the bottom area (9 and 10).

*B. Behavior*

FM can also serve to model the dynamic behavior of a system by identifying *events,* hence allowing control of the execution sequence of these events. An event in FM is specified by its spatial *area or subgraph*, its *time*, the *event's own stages*, and possibly by other descriptors such as intensity or extent (strength). Fig. 5 shows the (non-basic) event *A car is received in the area before the arm* in the example.

In this example, for simplicity, an event is represented by only its region of occurrence in the FM diagram. Accordingly, to build a description of the behavior of the system, four events are identified, shown in Fig. 6.

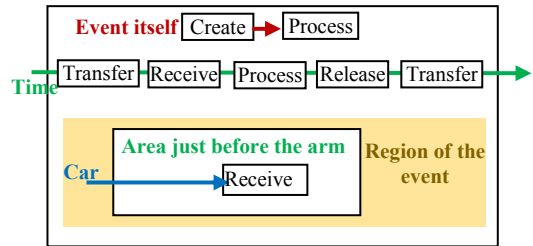

Fig. 5. The event *A car is received in the area just before the arm.*

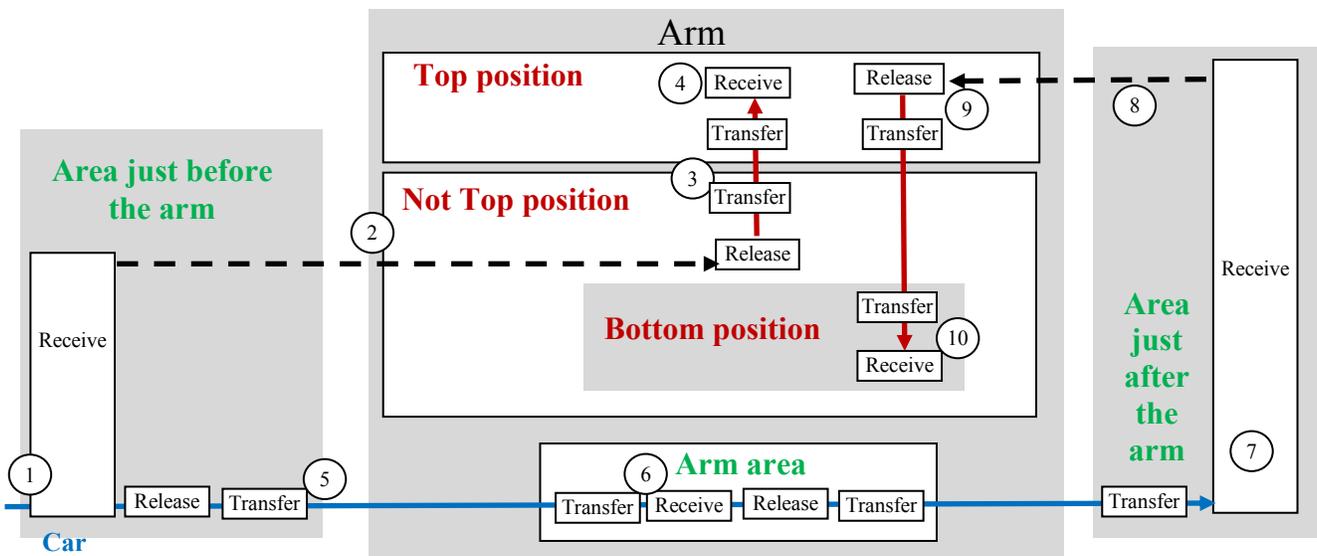

Fig. 4. FM representation of the implementation of a parking gate controller

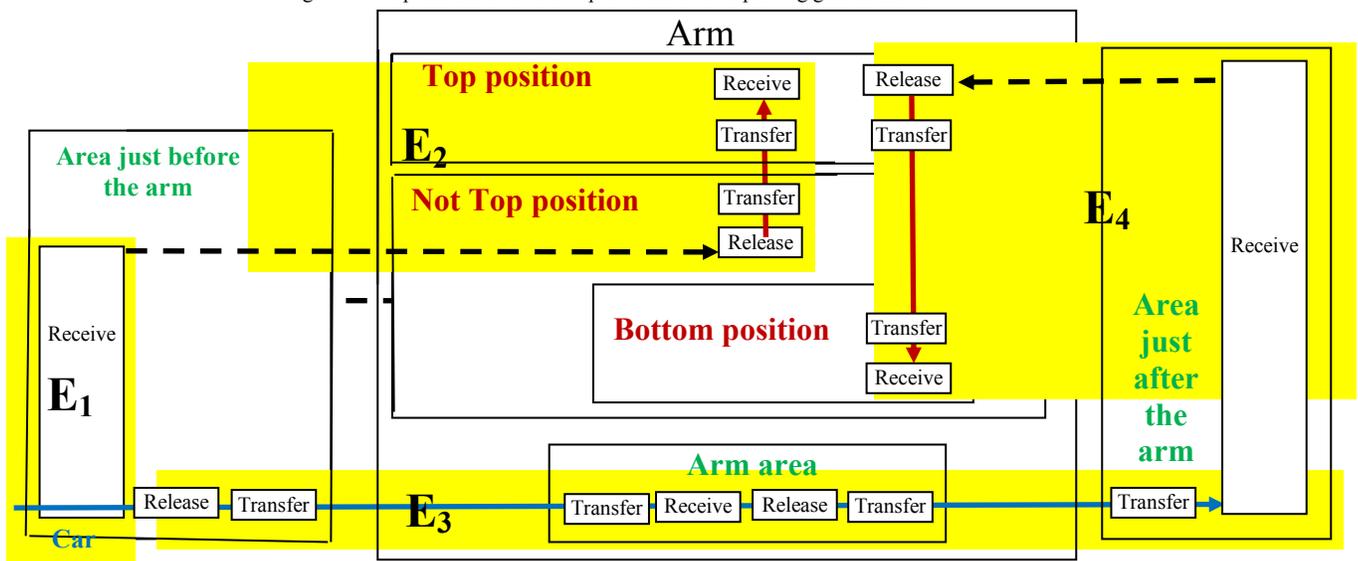

Fig. 6. Events

**Event 1 ($E_1$):** A car is received in the area before the arm.
**Event 2 ($E_2$):** The arm moves to the top position.
**Event 3 ($E_3$):** The car is received in the area after the arm.
**Event 4 ($E_4$):** The arm moves to the bottom position.

Fig. 7 shows execution control of the events. It is not difficult to write such a control in language such as,

*E1*
*If (If Arm is not in bottom) then E2*
*E3*
*E4*

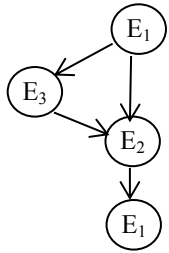

Fig. 7. Event execution control

## IV. APPLYING FM TO PETRI NETS

This section applies FM to some Petri net examples. The purpose is to explore the relationship between the two techniques and to demonstrate the benefits of FM for understanding of Petri nets.

### A. Elementary Nets

Desel and Reisig [14] provide the diagram shown in Fig. 8 as an example of elementary nets. The figure models the control part of a vending machine. At the initial state the machine is waiting for a coin to be inserted. An inserted coin is either rejected or accepted, depending on a check to determine whether it is "not part of the system model." If the coin is rejected, the system returns to its initial state. Otherwise the system first dispenses an item and then returns to its initial state [14]. Fig. 9 shows the corresponding FM representation. A state is a *thing*, thus, it is assumed it is ON (circle 1; e.g., power supply).

A coin is inserted, received (2-3), and stored in the machine. It is processed (4); if rejected, then it is released (5). If accepted, then this triggers (6) release of the product (7).

Consider the "events" of the Petri nets shown in Fig. 8 in terms of the events shown in Fig. 10:
**Event 1 ($E_1$):** Ready for insertion
**Event 2 ($E_2$):** Coin is inserted
**Event 4 ($E_3$):** Reject coin
**Event 5 ($E_4$):** Accept coin
**Event 6 ($E_5$):** Dispense item
Figure 11 shows the event execution control.

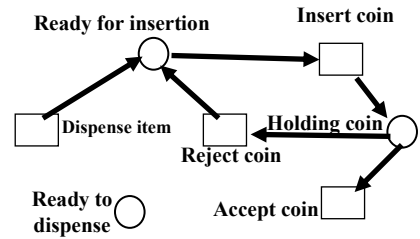

Fig. 8. Elementary net representing the control structure of a vending machine (redrawn, partial from [14]).

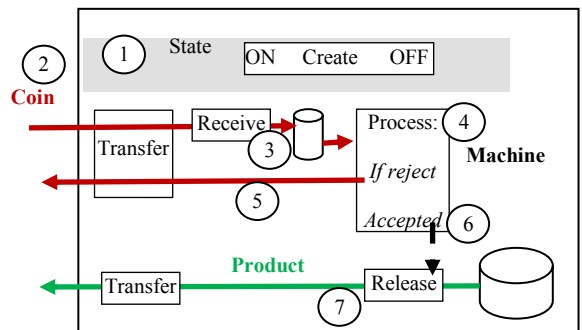

Fig. 9. FM representation of the example

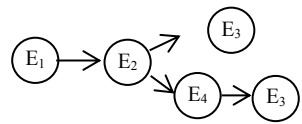

Fig. 11. Execution control

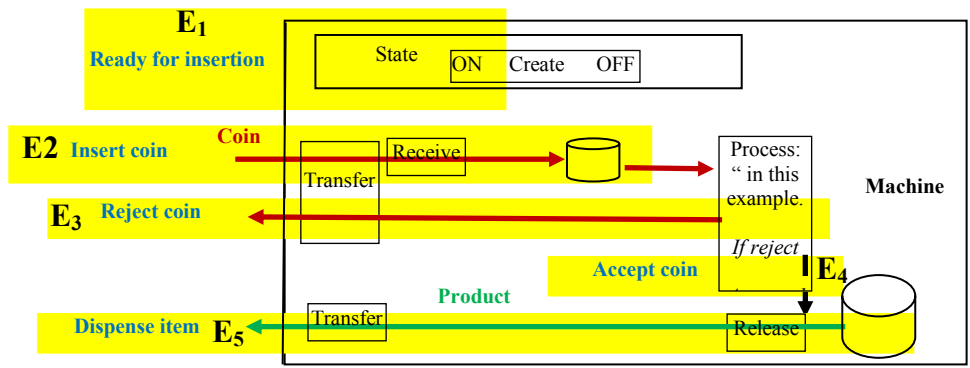

Fig. 10. Sequence of events from an FM perspective

Note that *Receiving* a coin implies *holding it* in the Petri net diagram. Also, the Petri net notion of "token" seems to be applied to two types of things in FM: the coin, and the items, whereas, according to Desel and Reisig [14] there is "only one token in the net" in this example.

### B. Vending Machine Modeling

According to Spiteri Staines [15], Petri nets are classifiable into four main categories: (i) elementary nets, (ii) normal Petri nets, (iii) higher order nets, and (iv) timed Petri nets or Petri nets with time. Each category has a specific use for systems engineering and software engineering; thus they can clearly assist with issues in requirements engineering.

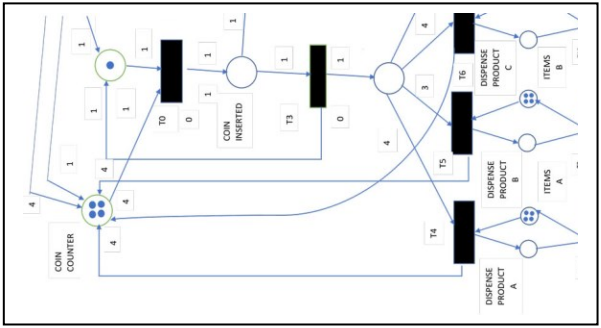

Fig. 12. Petri net representation (redrawn, partial from [15])

Spiteri Staines [15] gives examples of these categories in terms of a case study of a vending machine (see sample representation in Fig. 12). The main steps in the operation of the vending machine can be summarized as follows: (i) Coin insertion, (ii) Item selection and dispersal, and (iii) Refill of item.

The same problem (with some simplification, no refill) can be modeled in FM model as shown in Fig. 13. A person inserts a coin (circle 1) that flows to the machine to be processed (2) and stored in coin storage (3). Processing the coins also triggers the creation of data values (4) that are added to a total of coin values that flow to be compared after an item is selected (5). The machine continues to accept coins until an item is selected (6), when the selection flows to the machine to be processed (7). This triggers the retrieval of a price (8). Identifying the price along with the quantity input by the parson (9) creates the total in (10). The total flows to be compared with the value of entered coins. If the value is greater or equal to the total, the extra coins (if any) are calculated and processed (12), triggering the coin storage to release any extra coins to the customer (13). Otherwise (coin value = total), which triggers releasing the product (14) to the customer.

Fig. 14 models the dynamic behavior of the ATM through identifying *events* as follows.

**Event 1 ($E_1$):** The ATM receives coins.
**Event 2 ($E_2$):** The coins are processed and stored.

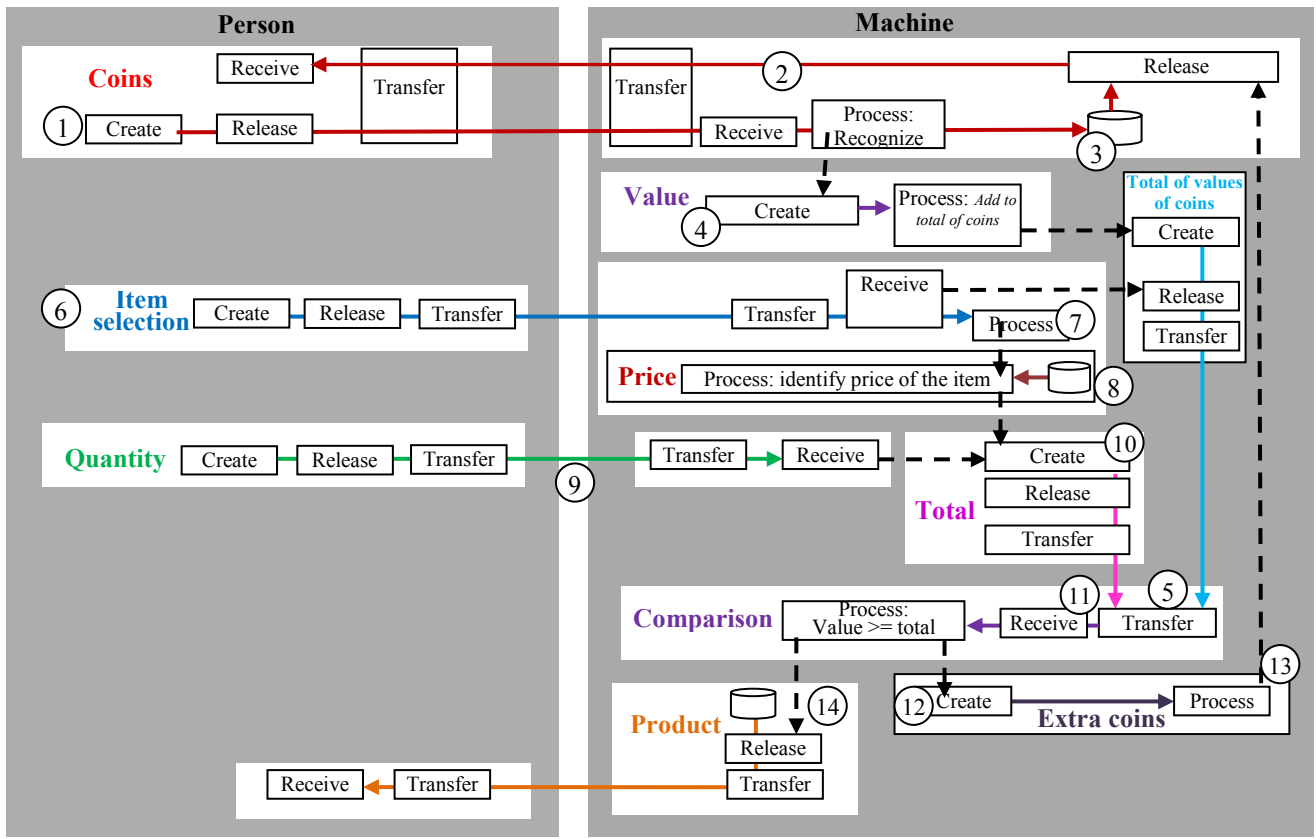

Fig. 13. FM representation of the vending machine.

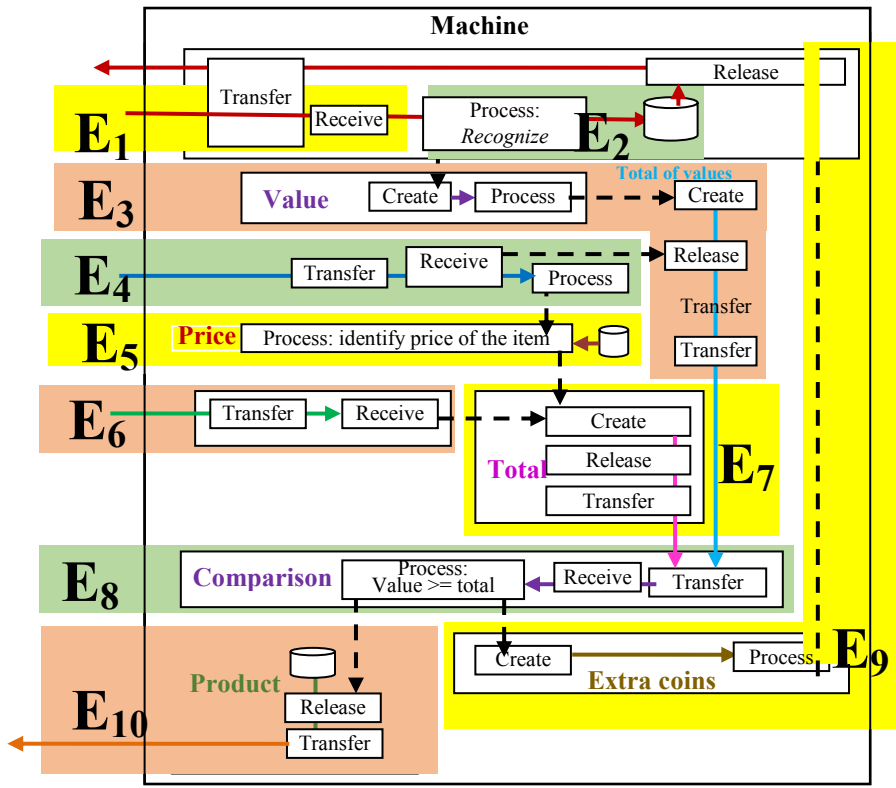

Fig. 14. Events.

**Event 3 ($E_3$):** The coins are processed and a value is sent for comparison.
**Event 4 ($E_4$):** The item selection is received, and
**Event 5 ($E_5$):** The price is retrieved.
**Event 6 ($E_6$):** The quantity is received.
**Event 7 ($E_7$):** The total price is calculated from the item price and the quantity.
**Event 8 ($E_8$):** The total price and the coin value are compared
**Event 9 ($E_9$):** It is found that the value is greater than the total price; thus the extra coins are identified and released to the customer.
**Event 9 ($E_9$):** The total price and the coin value are compared and it is found that the value is equal to the price.
**Event 10 ($E_{10}$):** The product is released to the customer.

Fig. 15 shows the sequence of events.

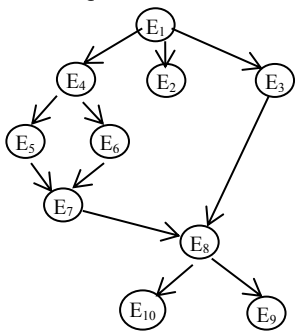

Fig. 15. Execution control

### C. Vending machine with time

According to Spiteri Staines [15], the timed Petri net adds time to transitions, places, or arcs. The corresponding diagram is almost identical to the vending machine shown in a general Petri net diagram. The only changes are the transition types, where the immediate transitions are converted to timed transitions [15]. The transition insert coin has a time value of 100, which could represent the time in seconds needed by a customer to insert a coin. Accordingly, because of space limitations, we will not develop the whole FM diagram. To illustrate how to model time, we focus on two instances of timing constraints.

1, The customer is given 100 seconds to insert the first coin after the start of the transaction.

2. The customer is given 60 seconds to insert the next coins, otherwise the coins are returned to the customer.

As seen in Fig. 16, we assume that the machine is originally OFF and a start signal is received by the machine (1) which turns it ON (2), and this triggers clock 1 to the ON state (3).

Then, **only for the first coin**:
- The ON state triggers the machine *clock 1* to create a timing period of 100 seconds (4) to allow the customer to insert coins.
- If *time* > 100 (5) without coins inserted, the machine turns OFF again (6).
- If the machine *receives* (7) a coin within the 100 seconds, then,

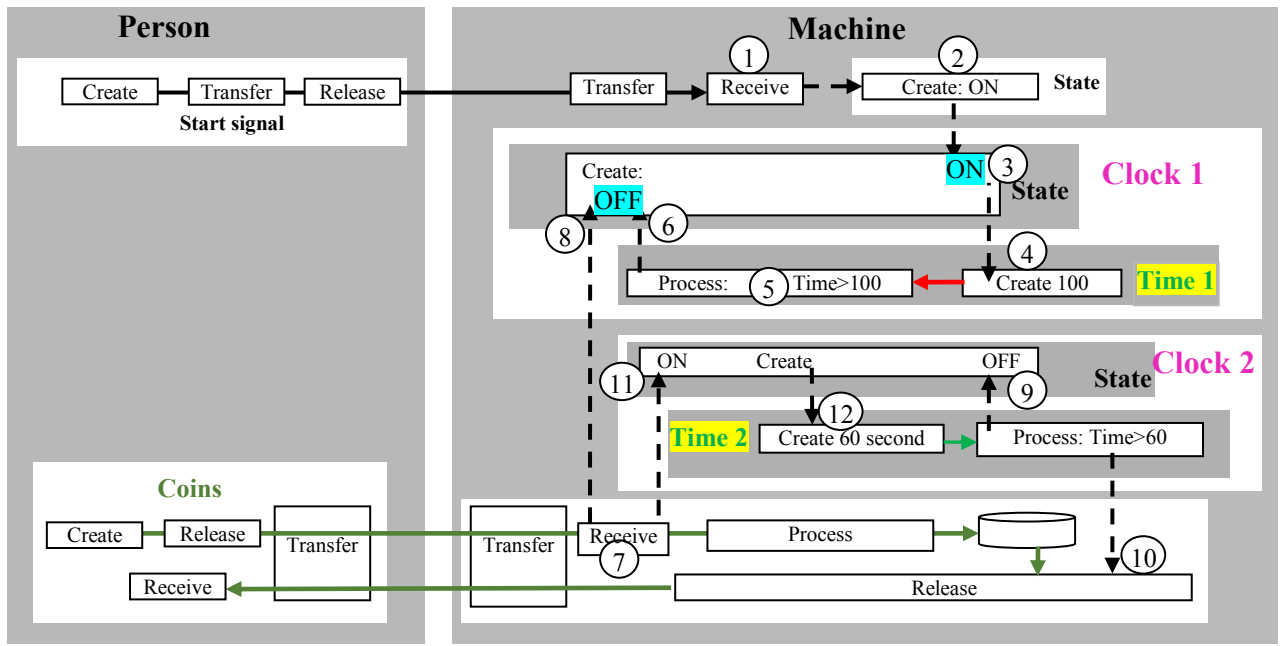

Fig. 16. Examples of two timing constraints

(i) It triggers *clock 1* to the OFF state (8).
(ii) It turns ON *clock 2* (9) and sets it to 60 seconds for the next coin (10).
(iii) It processes the coin, transfers it to storage and finds its value to be sent for comparison.

**For any next coin**:
- If time is greater than 60 seconds, it turns *clock 2* OFF (9) and releases the coins back to the customer (10).
- If the next coin is received within 60 seconds,
   (i) It triggers *clock 2* ON again (11) which initializes *clock 2* to another 60 seconds (12).
   (ii) Additionally, it processes the coin, transfers it to storage and finds its value to be sent for comparison.

Handling each coin in turn continues in this manner until the sequence is interrupted by selection of an item, which triggers clock 2 to the OFF state.

Events can be identified as follows (see Fig. 17):

**Event 1 ($E_1$)**: A start signal is received.
**Event 2 ($E_2$)**: Clock turns ON and time is set to 100 seconds.
**Event 3 ($E_3$)**: Time > 100 seconds; thus, clock 1 is turned OFF.
**Event 4 ($E_4$)**: A coin is received.
**Event 5 ($E_5$)**: Clock 1 is turned OFF.
**Event 6 ($E_6$)**: Clock 2 is turned ON, thus initializing time of clock 2 to 60 seconds.
**Event 7 ($E_7$):** Time of clock 2 > 60 seconds. Thus clock 2 is turned OFF and coins are released back to the user.
**Event 8 ($E_8$)**: A next coin is inserted within 60 seconds; thus, it turns clock 2 OFF, causing it to be re-initialized to 60 seconds. Also, the value is found and added to the current total value of coins.

Note that turning clock 1 OFF again does not affect anything because it is already OFF.

**Event 9 ($E_9$)**: The coin is processed and stored, in addition to triggering a determination of its value that is added to the total value.

Fig. 19 shows the execution control where the last arrow in the figure leads to the rest of the model where receiving coins is ended by selecting an item.

## V. CONCLUSION

This paper has applied the FM modeling language to several Petri net examples to examine their representations. The results point to the viability of the approach for exploring the underlying assumptions in Petri nets. It seems that FM can express all types of Petri nets such as regular and timed nets and can be used as an alternative way to explain Petri nets along with English text and ad hoc graphs. Other benefits of importing the FM language into the field of Petri nets seem plausible, however, this preliminary paper is not conclusive. Further research is needed to investigate this issue in more depth. This includes the impressibility of the FM language. Can Petri nets provide a formal base for FM? How can the dynamism of Petri nets be imported into FM?

Note that it can be claimed that the complexity of FM diagrams may present difficulties; however, solutions to visual complexity have already been implemented in many engineering systems (e.g., aircraft and high-rise building schemata) through multilevel simplifications. The details can be lumped together by omitting stages and unifying flows in the model. FM diagrams can be simplified by removing the detailed stages. A more elaborate simplification can be produced by preserving the stages Create and Process under the assumption that the stages Release, Transfer, and Receive are implied by arrows. These levels of simplification are based on the underlying FM schema which remains the reference for such purposes.

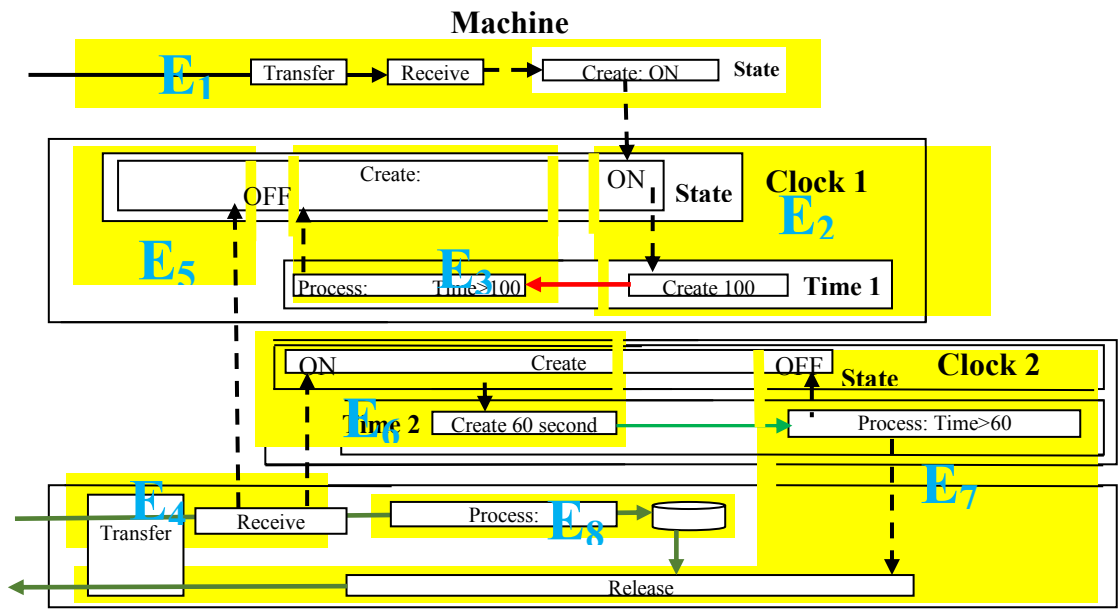

Fig. 17. The events.

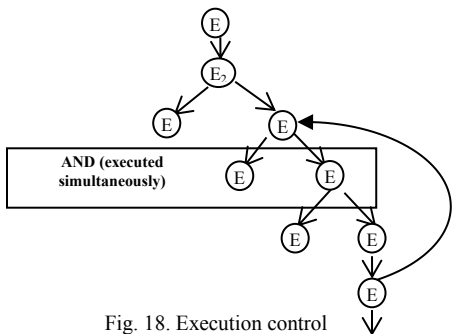

Fig. 18. Execution control